# INTEGRATED GEOPHYSICAL MEASUREMENTS ON A TEST SITE FOR DETECTION OF BURIED STEEL DRUMS

*Marco Marchetti ([1]) and Alessandro Settimi ([1])*

([1]) Istituto Nazionale di Geofisica e Vulcanologia (INGV), Via di Vigna Murata 605, I-00143 Rome, Italy

*Corresponding author: Marco Marchetti

Istituto Nazionale di Geofisica e Vulcanologia (INGV)

Via di Vigna Murata 605

I-00143 Rome

Italy

Tel: +39-06-51860300

Fax: +39-06-51860397

Email: marco.marchetti@ingv.it

**Abstract**

Geophysical methods are increasingly used to detect and locate illegal waste disposal and buried toxic steel drums. This study describes the results of a test carried out in clayey-sandy ground where 12 empty steel drums had previously been buried at 4-5 m below ground level. This test was carried out with three geophysical methods for steel-drum detection: a magnetometric survey, electrical resistivity tomography with different arrays, and a multifrequency frequency-domain electromagnetic induction survey. The data show that as partially expected, the magnetometric and electromagnetic induction surveys detected the actual steel drums buried in the subsurface, while the electrical resistivity tomography mainly detected the changes in some of the physical properties of the terrain connected with the digging operations, rather than the actual presence of the steel drums.

# 1. Introduction.

Due to the recent advances in electronics and in data-processing software, and to the increased experience in data interpretation, many cases of buried illegal waste have been discovered through the use of geophysical surveys. Furthermore, the low cost of obtaining the geophysical data and their characteristic noninvasive techniques have promoted a great increase in their use in the territory.

Even if the magnetometric method is used more frequently, other geophysical techniques can be used in such investigations (Emerson et al., 1992; Pierce and De Reamer, 1993; Foley, 1994; Vogelsang, 1994; Dahlin and Jeppsson, 1995; Daniels et al., 1995; Bernstone et al., 1996; Gibson et al., 1996; Huang and Keiswetter, 1997; Godio et al., 1999; Orlando and Marchesi, 2001; Marchetti et al., 2002; Chianese et al., 2006; Ting-Nien and Yi-Chu, 2006; Hamzah et al., 2009; Ruffell and Kulessa, 2009). Indeed, the choice of the methodology to be used will depend on the physic characteristics of the materials and the depth of the targets.

This study describes the results from a test site where several integrated geophysical methods were used to detect some buried steel drums (Morucci, 2003). A 5-m-deep, 3-m-wide and about 10-m-long hole was dug into the slope of a valley that is characterized by clayey-sandy deposits (Figure 1). The site is located about 50 km from Rome. Twelve empty steel drums were buried in a vertical orientation inside the hole, with their top surface at a depth of 4 m to 5 m below ground level, to simulate a genuine case of hidden drums that might contain, for example, toxic waste. The longer side of the hole was in an east-northeast to west-southwest orientation, and each drum was 0.88 m high with a diameter of 0.58 m.

Three types of surveys were carried out in this test area: a magnetometric survey, electrical resistivity tomography (ERT) with different arrays, and a multifrequency frequency-domain electromagnetic (FDEM) induction survey. While magnetometer and induction surveys are regularly used for the detection of buried drums and tanks, ERT is more widely used in studies of groundwater pollution, to determine the presence of leachates in landfills, and in the study of the possible escape of leachates from municipal waste.

In the present study, the ERT measurements were carried out to determine whether the metal drums could induce resistivity variations in the data. The *in-situ* sediments have low resistivity values (around tens of ohm·m), which are similar to those that are likely to be found in urban waste dumps, where the magnetic induction method might not provide clear answers because of the high levels of iron masses scattered in such waste.

In practice, the goal was to determine whether steel drums buried in a landfill site of municipal waste can be detected with these geophysical techniques (Marchetti et al., 1995; Marchetti, 1997, 2000).

**2. Magnetic measurements.**

The magnetometric technique is the geophysical method that is most frequently used for environmental problems (Bevan., 1983; Tyagi et al., 1983a, 1983b; Barrows and Rocchio, 1990; Roberts et al., 1990; Schlinger, 1990; Gilkenson et al., 1992; Foley, 1994; Cochran and Dalton, 1995; Gibson et al., 1996; Ravat, 1996; Marchetti et al., 1998; Eskola et al., 1999; Godio, 2000; Furness, 2001, 2002, 2007; Marchetti et al., 2002; Sheinker et al., 2009). As such, we can say that among the potential techniques for geophysical exploration of the subsoil, magnetometry generally appears to be one of the most effective, rapid and precise for the location of buried ferromagnetic masses (Marchetti, 1997, 2000; Marchetti and Meloni, 1997). Magnetometric surveys allow the detection of the surface effects and the local disturbances in the Earth magnetic field that are generated by buried ferromagnetic objects. These effects are known as magnetic anomalies, and can result from the combination of the Earth magnetic field with the induced and permanent magnetization of the magnetic targets. Natural bodies (such as a magnetic ore deposit) and man-made iron and steel objects (such as pipelines, vehicles, rails, mines and, as in our case, buried drums) can produce local deformations in the geomagnetic field. The detectability of magnetic objects by a magnetometer depends on their effective magnetic mass, the intensity of the magnetization, and the distance from the magnetometer. The intensity of the anomalies varies inversely as the square (for a monopole) or the cube (for a dipole) of the distance (Breiner, 1973).

On this test site, the survey was carried out along 12 profiles, each spaced 2 m apart, with a sampling rate of every 1 m. Around the buried drums, an area of 720 m$^2$ was covered with about 360 measurements. The magnetic data were collected using an optical pumped cesium magnetometer, the Geometrics model G-858, in gradiometer configuration: two sensors were mounted on a vertical staff at a distance of 1 m and 1.5 m from ground level. A magnetic base station with sampling rate of 1 s was used during the data collection, and the measurements were corrected for the magnetic diurnal variation.

Figure 2a shows the map of the anomalies of the total intensity of the Earth magnetic field related to the top sensor measurements. This map shows a typical dipolar magnetic anomaly that is characterized by a well-defined maximum and a less-intense minimum. This anomaly is clearly connected to the buried steel drums, and it reaches a total intensity of about 290 nT, with its main axis in a north-south orientation. The signature of this anomaly is similar to that obtained on another test site by Marchetti et al. (1998). The broad minimum appearing in the left upper quadrant of this map is related to the presence of some wire netting. Figure 2b shows the map of the vertical magnetic gradient, calculated starting from the data collected by the two cesium sensors. The vertical gradient characterizes the steel-drum anomaly more precisely, as it can detect shallow buried targets better than the total intensity magnetic field (Breiner, 1973). In the ferromagnetic objects, induced and remnant magnetization contribute to the production of a single magnetic anomaly. In these cases, the remnant magnetization can be much larger than the value of the induced magnetization (Ravat, 1996). The assemblage of steel drums can be viewed as the combination of single individual permanent magnetizations that partly compensate for each other, leaving almost only the induced part. A very large number of drums can completely cancel out the remnant magnetization contribution (Breiner, 1973; Marchetti et al., 1996, 1998). The main axis of the drum anomaly was north-south oriented, in agreement with the direction of the Earth magnetic field (induced magnetization).

**3. Geoelectrical measurements.**

The geoelectrical technique (ERT) is based on the analysis of the underground electric fields generated by a current flow injected from the surface. This resistivity method is based on the electric conduction in the ground, and it is governed by Ohm's law. From the current source $I$ and potential difference $\Delta V$ values, an apparent resistivity value $\rho_a$ can be calculated as $\rho_a = k\,(\Delta V/I)$, where $k$ is a geometric factor that depends on the arrangement of the four electrodes. A pair of electrodes (A, B) are used for the current injection, while potential difference measurements are made using a second pair of electrodes (M, N). The potential is then converted into apparent resistivity, and then by inversion to the true resistivity, which depends on several factors: mainly the lithology of the soil, and its porosity, and the saturation and conductivity of its water pores.

ERT is a powerful tool that is widely used for environmental site assessments and to map leachate concentrations within closed and unconfined landfill sites (Bernstone and Dahlin, 1988, 1997; Dahlin, 1996, 2001; Dahlin and Bernstone, 1997; Bernstone, 1998; Loke, 1999; Wisèn et al., 1999; Bernstone et al., 2000; Nasser et al., 2003; Lillo et al., 2009). In the ERT method, a multiple electrode string is placed on the surface, and then using computer-controlled data acquisition, each electrode can serve both as a source and as a receiver; thus a large amount of data can be collected quickly during a survey.

The surveyed depth depends on the length of the geoelectric extension and on the selected sequence of measurements. A numerical inversion routine is used to determine the probable electrical resistivity distribution of the subsurface. Due to the progress in both electronics and data-processing software, it is now possible to make real three-dimensional tomography images using direct-current measurements on electrode grids (Loke and Barker, 1996; Dahlin and Loke, 1997; Ogilvy et al., 1999, 2002; Finotti et al., 2004; Morelli et al., 2004; Fischanger et al., 2007).

A north-northwest to south-southeast oriented ERT line was carried out using a Syscal R2 resistivity meter equipped with a line of 48 electrodes (stainless steel stakes) spaced 1 m apart and connected through automatic switching to a three multinode box, each node of which can drive 16 electrodes. This profile was centered orthogonally on the drums, and the measurements were carried out using different array configurations: Wenner, dipole-dipole and pole-dipole. The Wenner array is an attractive choice for surveys carried out in areas with a lot of background noise (due to its high signal strength), and also when good vertical resolution is required. The dipole-dipole array might be a more suitable choice if good horizontal resolution and data coverage is important (assuming the resistivity meter is sufficiently sensitive and there is good ground contact). If a system has a limited number of electrodes, the pole-dipole array with measurements in both the forward and reverse directions would be a viable choice (Loke, 1999).

To determine the values of the ground resistivity, ERT with the Wenner array was performed in an area that was not affected by the excavation (about 9 m further downhill). The data analysis and modeling were carried out using a commercial geophysical inversion program (Res2dinv).

The profiles carried out in this geoelectrical survey are shown in Figure 3. In the ERT profile of Figure 3a, the resistivity values rise regularly from the shallower to the deeper terrain, according to soil moisture variations. This profile was performed away from the buried drums. Instead, for the experimental site, a large resistivity region is present that corresponds to the buried steel-drum cluster (Figure 3b). The increase in the resistivity acquired by the ground was caused by the digging operations and by terrain reworking effects, rather than by the conductivity of the steel drums. Therefore, the geoelectrical survey only detected the presence of the drums in the subsurface as an indirect effect. ERT performed with different electrode arrays detected the resistivity increases, although various images of this high resistivity zone are shown because of their different geometrical characteristics.

In general (Loke, 1999), the Wenner array is good for the resolving of vertical changes (i.e. horizontal structures), while it is relatively poor for the detection of horizontal changes (i.e. narrow vertical structures). Compared to the other arrays, the Wenner array has a moderate depth of investigation. Among the common arrays, the Wenner array has the strongest signal strength. This can be an important factor when a survey is carried in areas with high background noise. One disadvantage of this array for two-dimensional surveys is the relatively poor horizontal coverage as the electrode spacing is increased. This can be a problem if the system used has a relatively small number of electrodes.

The dipole-dipole array has been, and still is, widely used in resistivity and induced-polarization surveys, because of its low electromagnetic (EM) coupling between the current and potential circuits. The dipole-dipole array is very sensitive to horizontal changes in resistivity, although relatively insensitive to vertical changes in resistivity. This means that it is good for the mapping of vertical structures, such as dykes and cavities, but relatively poor for the mapping of horizontal structures, such as sills or sedimentary layers. In general, this array has a shallower depth of investigation compared to the Wenner array, although for two-dimensional surveys, this array has better horizontal data coverage than the Wenner array. This can be an important advantage when the number of nodes available with the multi-electrode system is small. One possible disadvantage of this array is the very small signal strength.

The pole-dipole array also has relatively good horizontal coverage, but it has a significantly higher signal strength compared with the dipole-dipole array. Unlike the other common arrays, the pole-dipole is an asymmetrical array. One method to eliminate the effects of this asymmetry is to repeat the measurements with the electrodes arranged in the reverse manner. However, these procedures will double the number of data points and consequently the survey time. Similar to the dipole-dipole array, this array is probably more sensitive to vertical structures. Due to its good horizontal coverage, this is an alternative array for multi-electrode resistivity meter systems with a relatively small number of nodes. The signal strength is lower compared with the Wenner array, but higher than the dipole-dipole array. In particular, Figure 3b-d shows these ERT measurements, respectively corresponding to the Wenner, dipole-dipole and pole-dipole configurations that were used on this test site.

The different array resolutions from the ground reworking are visible in the ERT sections. The pole-dipole array (carried out with 32 electrodes) appears to be the only one of these arrays that can more precisely detect the high resistivity zone, although it shows a slightly eccentric image, as it is an asymmetric array.

Resistivity measurements for mapping the geology of different terrains have been applied for more than half a century. However, some deficiencies have prevented this technique from being widely used for engineering aims. The first is that ordinary measurements of resistivity involve a relatively high number of performing operators, which is therefore expensive. Secondly, actual resistivity seldom has a diagnostic merit; it is just the lateral or vertical alterations in the resistivity that allow a physical interpretation.

## 4. Frequency-domain electromagnetic induction measurements.

The FDEM induction method for measuring ground resistivity, or more correctly, conductivity, is well known, and some extensive discussions of this technique can be found in the references given in the studies by McNeill (1980a, 1980b).

The FDEM induction method is based on the response of an induced alternating current in the ground. Consider a transmitter coil Tx energized with an alternating current at an audio frequency placed on the Earth (assumed to be uniform), and a receiver coil Rx located a short distance $s$ away. The time-varying magnetic field arising from the alternating current in the transmitter coil can induce very small currents in the Earth. These currents generate a secondary magnetic field $H_s$, which is sensed by the receiver coil, together with the primary field, $H_p$.

In general, this secondary magnetic field is a complicated function of the inter-coil spacing $s$, the operating frequency $f$, and the ground conductivity $\sigma$. Under certain constraints, which are technically defined as "operation at low values of induction number" (discussed in detail by McNeill, 1980a, 1980b), the secondary magnetic field is a very simple function of these variables. The ratio of the secondary to the primary magnetic field is linearly proportional to the terrain conductivity, a relationship that makes it possible to construct a direct-reading, linear-terrain conductivity meter by simply measuring this ratio. Given $H_s/H_p$, the apparent conductivity indicated by the instrument is defined by the equation: $\sigma_a = (4/\omega\mu_0 s^2)(H_s/H_p)$, where $\omega=2\pi f$ and $\mu_0$ are the permeabilities of free space. The MKS units of conductivity are the mho (Siemens) per m, or more conveniently, the millimho per m.

In physical terms, if a conductive medium is present within the ground, the magnetic component of the incident EM waves induces eddy currents (alternating currents) within the conductor. These eddy currents then generate their own secondary EM field, which can be detected by the receiver, together with the primary field that travels through the air; consequently, the overall response of the receiver is the combined effects of both the primary and the secondary fields. The degree to which these components differ reveals important information about the geometry, size and electrical properties of any sub-surface conductors (Reynolds, 1997).

EM induction methods use ground responses to the propagation of EM waves to detect electrical conductivity variations. Some environmental applications of these methods are, for example, the detection of landfills, unexploded ordnances, buried drums, trenches boundaries, and contaminant plumes (McNeill, 1980a, 1980b; Tyagi et al., 1983a, 1983b; McNeill, 1994, 1997; Jordant and Costantini, 1995; Won et al., 1996, 1997; Bernstone and Dahlin,1997; Witten et al., 1997; Wisèn et al., 1999; Huang and Won, 2000, 2003a, 2003b, 2003c, 2004; Norton and Won, 2001).

This FDEM survey was carried out using a GSSI GEM 300 instrument, which is suitable equipment for the simultaneous measuring of up to 16 user-defined frequencies between 330 Hz and 20,000 Hz. As it works in multifrequency mode, it is possible to obtain not only a detailed underground map, but also information at different depths; indeed, the penetration depth of the electromagnetic signal into the subsurface is inversely proportional to the frequency. The secondary field measured by the receiver coil of the FDEM sensor is divided into in-phase and quadrature components that are expressed as percentage intensities of the signals relative to the primary-field strength. Note that this instrument is no longer in production, as it has now been replaced by the GSSI Profile EMP 400.

This FDEM survey was carried out along profiles that were north-northwest to south-southeast oriented, 20-m long, and at a distance of 2 m apart, to cover an area of 400 m$^2$. The data were acquired along each profile in steps of 1 m, working continuously at the frequencies of 1,925, 2,675, 3,725, 5,125, 7,125, 9,875, 13,725 and 19,025 Hz. Figure 4a shows the in-phase component map at these selected frequencies. Here, a monopolar anomaly can be seen that shows up more and more clearly with the decrease in frequency, and so with the increase in the penetration depth of the EM signal; this anomaly is associated with the buried steel drums. Figure 4b shows the quadrature component maps that describe the terrain conductivity variations at the different frequencies. As can be seen, the in-phase response is more sensitive to the buried steel drums, relative to the quadrature component of the induced magnetic field, which is linearly related to the ground conductivity (McNeill, 1983, Dahlin and Jeppsson, 1995). In all of the maps shown in Figure 4b, a sharp distinction is seen between a resistive zone on the left side of each map – corresponding to the highest part of the study area – and a conductive zone on the right side of each map – corresponding to the lowest part of the test area. This conductivity/ resistivity variation is probably linked to the soil moisture variations that themselves are related to the valley slope, and the effects are most visible at high frequencies. The effects associated with the steel drums are also seen in these maps.

The monopolar signature of the electromagnetic anomaly allows the better locating of the underground position of the steel drums than the dipolar signature of the magnetic anomaly. As magnetic anomalies are inherently dipolar in nature, the target is thus commonly located at the slope, rather than at the peak, of an anomaly. In contrast, it can be shown theoretically by forward modeling that an EM anomaly is almost monopolar, and centered directly above the target; consequently, this is easier to interpret than dipolar magnetic anomalies (Won et al., 1996).

It is very interesting to compare the data obtained by the FDEM induction survey with those of the ERT: in the first case, the variations in the terrain conductivity and the buried steel drums were evident; in the second case, the measurements have only revealed the variations in the resistivity of the subsoil. The presence of the steel drums was not enough to lower the resistivity values in the subsoil. Resistivity measurements are possible if there is a resistivity contrast between the target and the hosting terrain. This also depends on the volume related to the depth and the electrode space. Instead, the EM measurements refer more to the absolute value of resistivity than to its contrast.

McNeill (1980a, 1980b) provided a deep discussion of the physical principles for the mapping of the electrical conductivity of the ground by applying magnetically induced currents at low frequencies. Our present study has confirmed that some benefits can be gained from working at low values of induction number. These benefits include fine conductivity resolution, considerable reduction in the manpower necessary to perform the conductivity measurements, and simplified calculation of the layered Earth response.

One point should be clearly underlined when these types of measurements are performed for the mapping of the geology of a terrain: these devices probe just the electrical conductivity. If the conductivity does not depend substantially on the geological environment, or even if other factors other than the geology affect the conductivity, the data from such measurements would be difficult to interpret and understand.

In our case here, the geoelectrical method (ERT) does not allow the discrimination of the clayey-sandy ground from the buried steel drums, as the conductivity measurements are performed using only the current flow. Instead, the FDEM method can identify an anomaly due to the conductivity, as this method is based on the interactions of an EM wave with the conductor body.

**5. Conclusions.**

In this study, we have investigated the responses of some geophysical techniques for the detection of buried steel drums. Twelve empty steel drums were buried in clayey-sandy ground to simulate the actual burying of potentially toxic waste. On this test site, we carried out a magnetometric survey, ERT with different arrays, and an FDEM induction survey .

The ERT measurements were realized to determine whether these metal drums can change the characteristics of the resistivity of the soil where they are buried, as this terrain has low resistivity values that will be similar to those that are likely to be found in an urban-waste dump.

On the basis of both the geological characteristics of the test area and the results obtained from the various surveys, we can see that the magnetometric and FDEM induction methods can detect the steel drums buried in the subsurface: indeed, the target was indicated by a dipolar anomaly in the former case, and by a monopolar anomaly in the latter case. These methods were also carried out in less time, and with fewer operators needed. The ERT only detected changes in some of the physical properties of the terrain, and in particular, an increase in the electrical resistivity. These changes are associated with the digging operations and/or the empty steel drums. The FDEM induction survey was probably the optimum survey type, as this gave the best results for the locating of the buried drums (in addition to the magnetic methods that are commonly used in such studies) and the detection of the soil-conductivity variations. At the same time, the varying of the frequencies can provide information on the target depth. However, the EM instruments with fixed intercoil spacing achieve a depth of investigation that is less than that obtained by the magnetic method.

**Acknowledgements.**

The authors would like to thank Dr. Massimo Chiappini who prepared the test site, Geostudi Astier S.r.l. for measurements by the GEM 300, and Dr. Andrea Morucci for the magnetic and geoelectrical measurements.

**References.**

**Figure captions.**

**Figure 1.** Photograph of the preparation of the steel-drum arrangement.

**Figure 2.** Magnetic anomaly maps of the total intensity field (a) and the vertical magnetic gradient (b).

**Figure 3.** Electrical resistivity tomography profiles of the nearby terrain (a) and the test site for the Wenner (b), dipole-dipole (c) and pole-dipole (d) arrays. The lateral section of the steel-drum arrangement is also shown (b-d).

**Figure 4.** The multifrequency frequency-domain electromagnetic induction survey in-phase (a) and quadrature (b) component anomaly maps, for the different frequencies used.

**Figure 1**

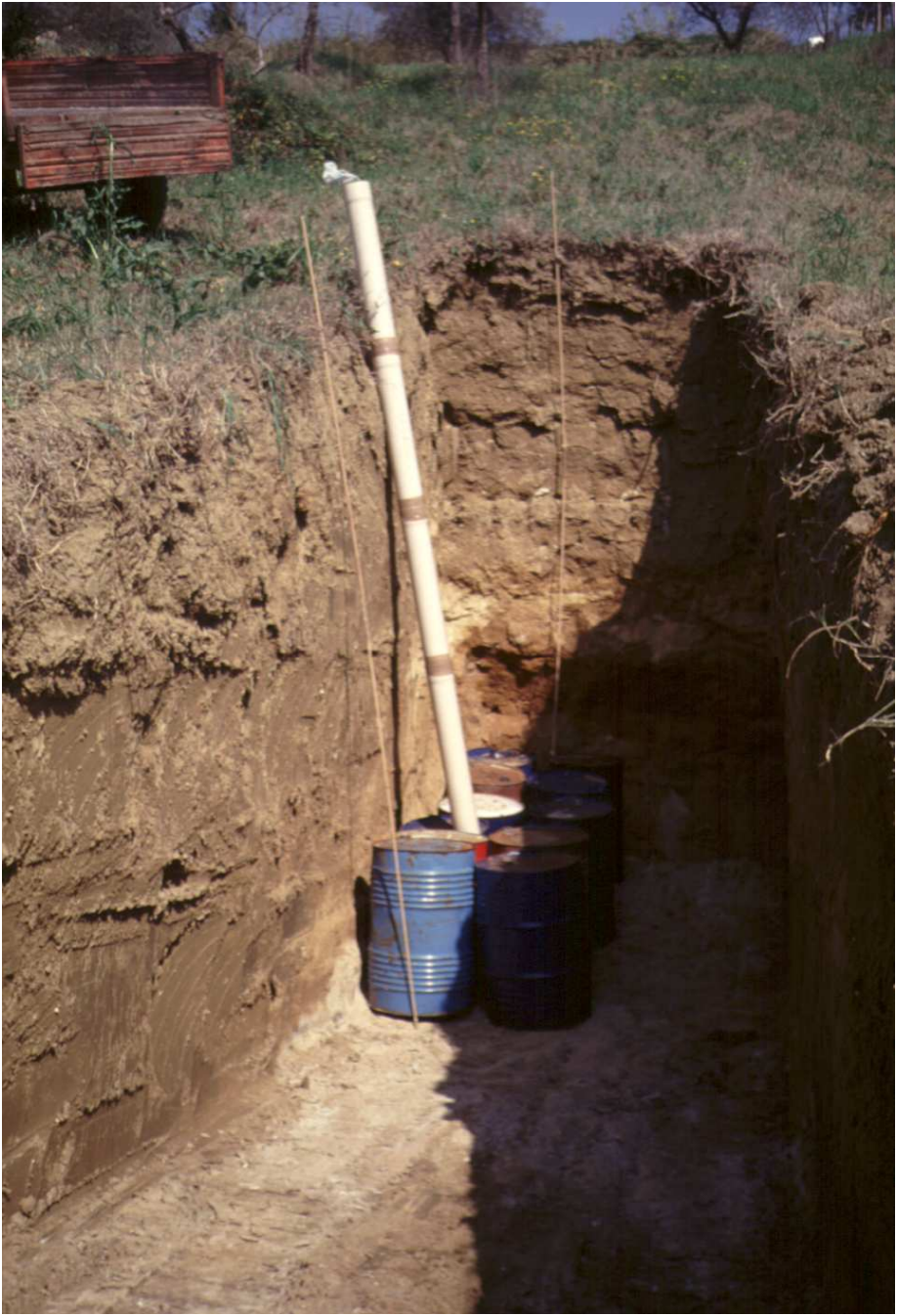

**Figure 2**

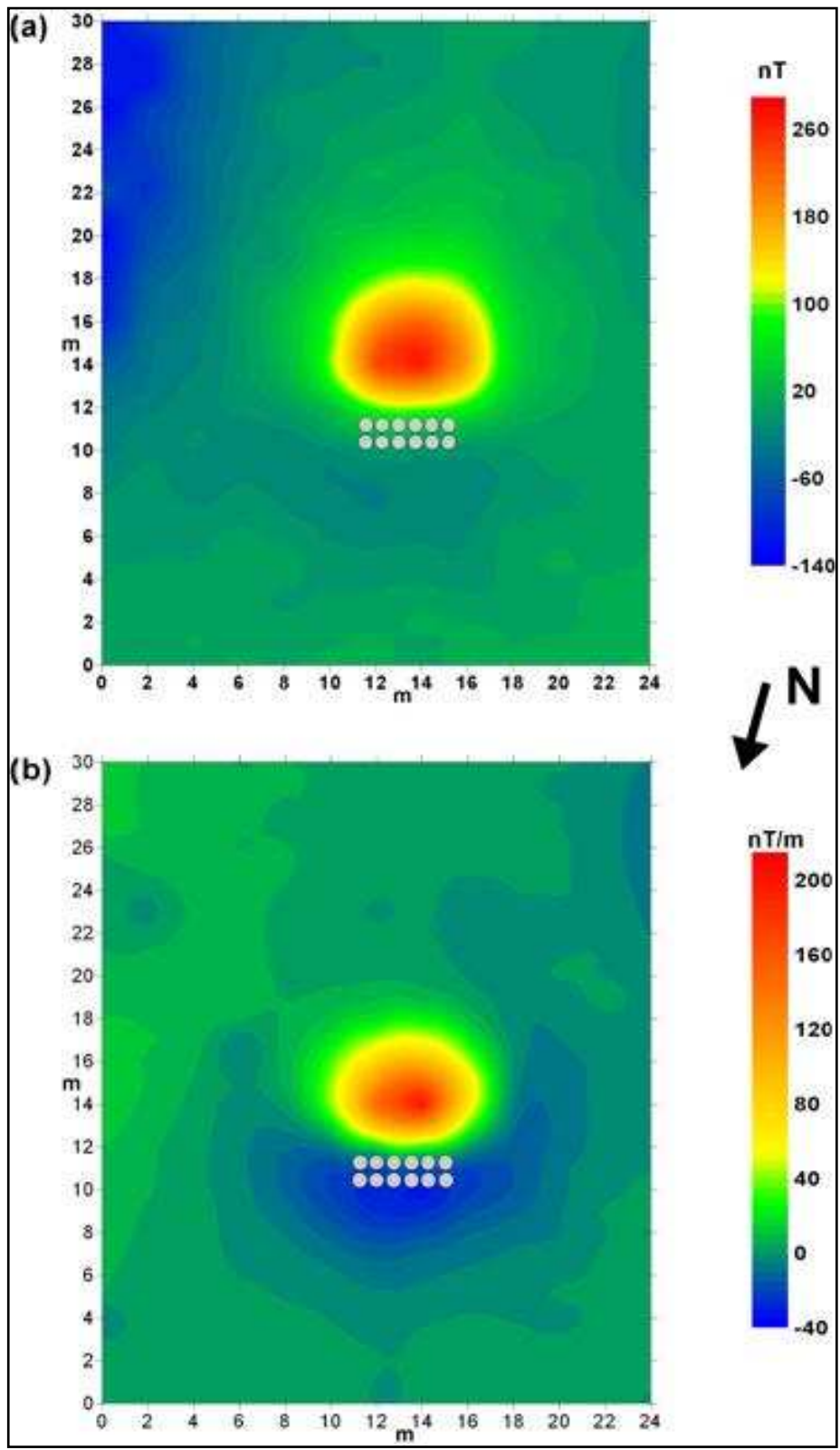

**Figure 3**

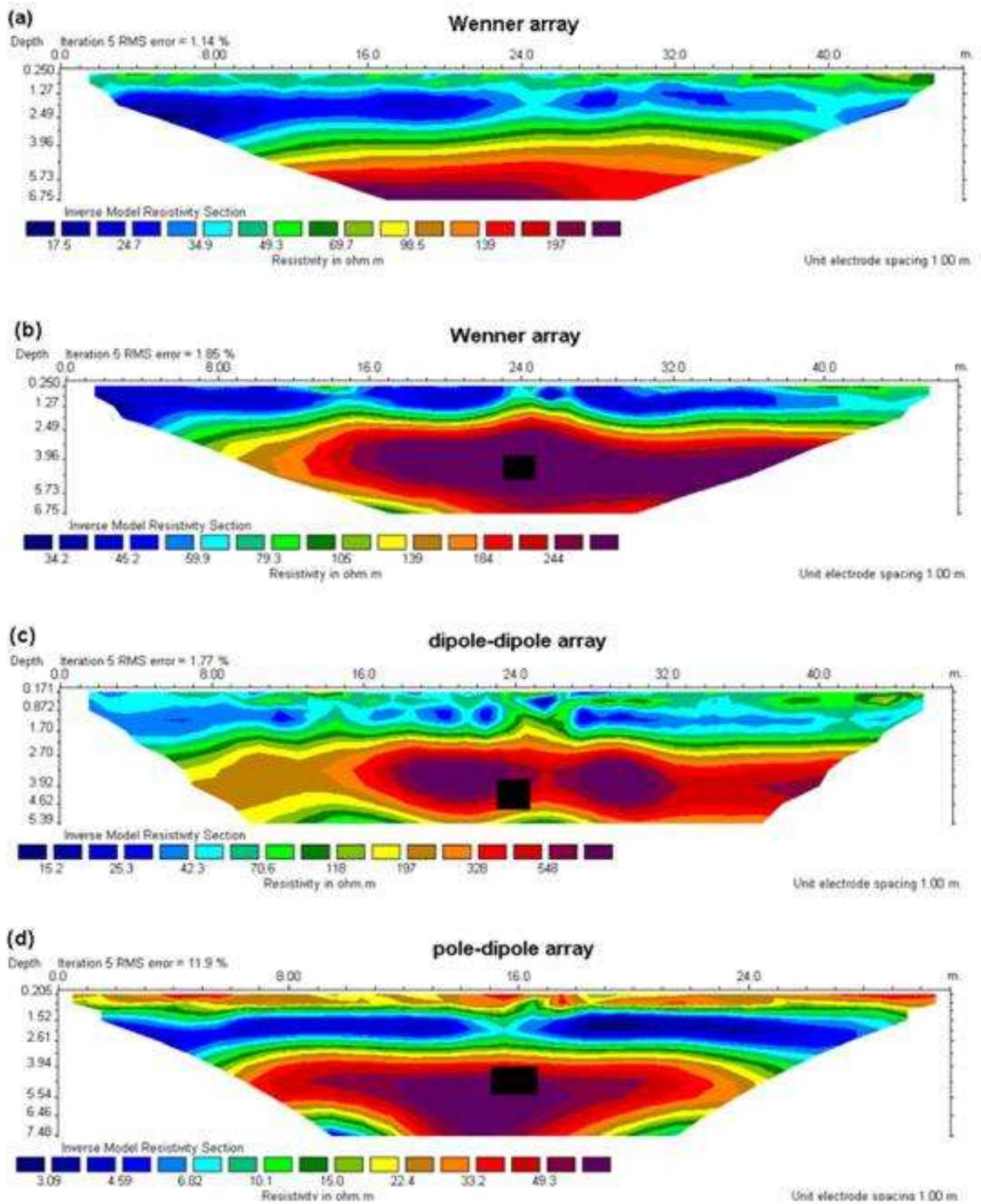

**Figure 4**

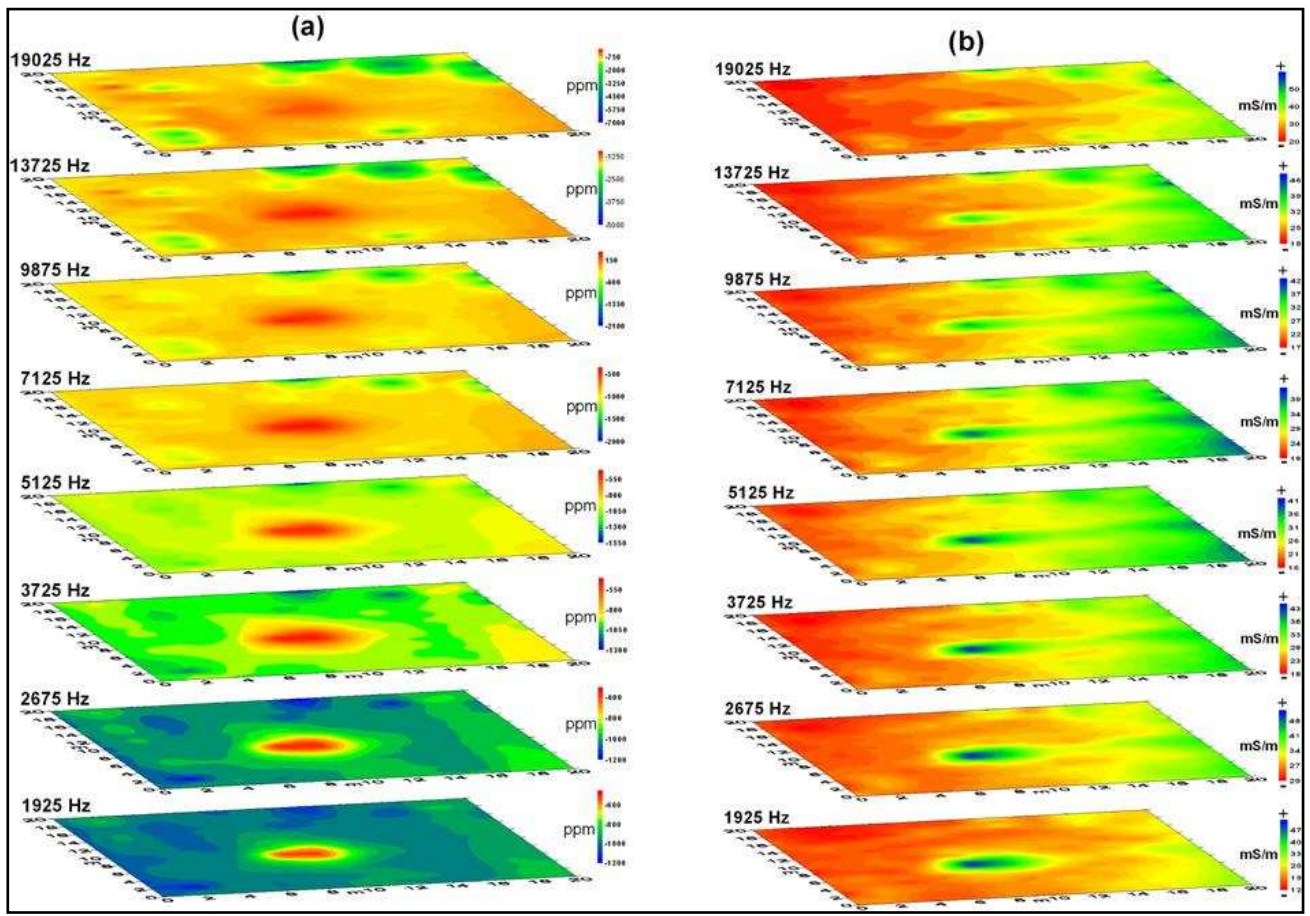